\documentclass{nature}

\usepackage{graphicx}
\usepackage{bm}
\usepackage{color}
\usepackage{soul}
\usepackage{lineno}

\title{Electronic structure and signature of Tomonaga-Luttinger liquid state in epitaxial CoSb$_{1-x}$ nanoribbons}

\author{Rui Lou$^{1,2,3,*}$, Minyinan Lei$^{1,3,}${\footnote{These authors contributed equally to this work.}} , Wenjun Ding$^{4}$, Wentao Yang$^{1,3}$, Xiaoyang Chen$^{1,3}$, Ran Tao$^{1,3}$, Shuyue Ding$^{1,3}$, Xiaoping Shen$^{1,3}$, Yajun Yan$^{3,5}$, Ping Cui$^{4,6}$, Haichao Xu$^{1,3,7,}${\footnote{xuhaichao@fudan.edu.cn}} , Rui Peng$^{1,3,7}$, Tong Zhang$^{1,3,7}$, Zhenyu Zhang$^{4}$ \& Donglai Feng$^{3,5,7,}${\footnote{dlfeng@ustc.edu.cn}}}

\begin{document}


\maketitle

\begin{affiliations}
 \item State Key Laboratory of Surface Physics, Department of Physics, and Laboratory of Advanced Materials, Fudan University, Shanghai 200438, China
 \item School of Physical Science and Technology, Lanzhou University, Lanzhou 730000, China
 \item Collaborative Innovation Center of Advanced Microstructures, Nanjing 210093, China
 \item International Center for Quantum Design of Functional Materials (ICQD), Hefei National Laboratory for Physical Sciences at Microscale, and Synergetic Innovation Center of Quantum Information and Quantum Physics, University of Science and Technology of China, Hefei 230026, China
 \item Hefei National Laboratory for Physical Science at Microscale, CAS Center for Excellence in Quantum Information and Quantum Physics, and Department of Physics, University of Science and Technology of China, Hefei 230026, China
 \item Key Laboratory of Strongly-Coupled Quantum Matter Physics, Chinese Academy of Sciences, School of Physical Sciences, University of Science and Technology of China, Hefei 230026, China
 \item Shanghai Research Center for Quantum Sciences, Shanghai 201315, China
\end{affiliations}

\newpage
\begin{abstract}
  Recently, monolayer CoSb/SrTiO$_3$ has been proposed as a candidate harboring interfacial superconductivity
  in analogy with monolayer FeSe/SrTiO$_3$. Experimentally, while the CoSb-based compounds manifesting as nanowires and thin films have been
  realized on SrTiO$_3$ substrates, serving as a rich playground, their electronic structures are still unknown and yet to be resolved. Here,
  we have fabricated CoSb$_{1-x}$ nanoribbons with quasi-one-dimensional stripes on SrTiO$_3$(001) substrates using molecular beam epitaxy,
  and investigated the electronic structure by \emph{in situ} angle-resolved photoemission spectroscopy. Straight Fermi surfaces without
  lateral dispersions are observed.
  CoSb$_{1-x}$/SrTiO$_3$ is slightly hole doped, where the interfacial charge transfer is opposite to that in monolayer FeSe/SrTiO$_3$.
  The spectral weight near Fermi level exhibits power-law-like suppression and obeys a universal temperature scaling, serving as the
  signature of Tomonaga-Luttinger liquid (TLL) state. The obtained TLL parameter of $\sim$0.21 shows the underlying strong correlations.
  Our results not only suggest CoSb$_{1-x}$ nanoribbon as a representative TLL system, but also provide clues for further investigations
  on the CoSb-related interface.
\end{abstract}

\maketitle

\section*{INTRODUCTION}
Over the past decade, substantial works have been performed on monolayer FeSe/SrTiO$_3$\cite{QWang2012,SHe2013,STan2013}, where the
cross-interface charge transfer and electron-phonon coupling are suggested to play important roles in enhancing the superconductivity\cite{JLee2014,
YXiang2012,ZLi2016,ZLi2019,GGong2019,QSong2019,BangJ2013,CaoHY2014}. Exploring similar systems could offer alternative
research aspects in understanding the interfacial interactions and to facilitate establishing a general picture. Such efforts in
cobalt-based compounds have led to a promising candidate, monolayer CoSb/SrTiO$_3$, which is predicted to adopt a tetragonal phase
and similar band structures as monolayer FeSe/SrTiO$_3$\cite{WDing2020}. Recently, a combined scanning tunneling microscopy/spectroscopy
(STM/STS) and magnetization measurement reports possible superconductivity in two-dimensional (2D) CoSb with orthorhombic lattice\cite{CDing2019},
where the triplet pairing symmetry is proposed\cite{CoSbTriplet}. Therefore, it is necessary to further explore
the electronic properties of CoSb-based interface.

Depending on different growth conditions using molecular beam epitaxy (MBE), the CoSb-based interface is shown to be a versatile system not
only manifesting as 2D thin film but nanowire with quasi-one-dimensional (Q1D) stripes, where a gap opening at Fermi level ($E_{\rm F}$)
is also observed\cite{CDing2019}. It thus serves as a rich playground for studying the electronic properties in different dimensions. When
the motion of electrons is confined in 1D, the Landau Fermi liquid theory tends to break down because the individual excitation is replaced
by the collective mode, which is highly correlated. It leads to the Tomonaga-Luttinger liquid (TLL) model that has succeeded in understanding
the distinctive characters of 1D systems, like the power-law-like behavior of various physical properties\cite{Tomonaga1950,Luttinger1963,
Dardel1991,Giamarchi2004,Voit1995}. In order to understand the electronic properties under dimensional confinement in CoSb-related nanowires,
it is critical to experimentally reveal their electronic structure by angle-resolved photoemission spectroscopy (ARPES).

In this paper, we have fabricated CoSb$_{1-x}$ nanoribbons on SrTiO$_3$(001) substrates utilizing MBE\cite{Composition}. The Q1D stripes on
nanoribbons give rise to straight Fermi surfaces (FSs). CoSb$_{1-x}$/SrTiO$_3$ is slightly hole doped, where the interfacial charge transfer
is opposite to that in monolayer FeSe/SrTiO$_3$. The near-$E_{\rm F}$ spectral weight exhibits power-law-like suppression and obeys a universal
temperature scaling, demonstrating the underlying TLL nature. The determined TLL parameter of $\sim$0.21 reveals that CoSb$_{1-x}$ is in the
strongly correlated regime. These results establish a model TLL system and provide clues for further studies on the CoSb-related interface.

\section*{RESULTS}
\textbf{Surface and structural characterization of as-grown nanoribbons}

\noindent The surface and structural characterization of CoSb$_{1-x}$ are presented in Fig. 1. Figure 1a-c displays the STM topographic images of Sample
I-III, respectively. The surface morphology shows nanoribbon feature with a typical step height of $\sim$1.2 nm (see Supplementary Fig. 1). Statistics over 35 nanoribbons in Sample III gives the ribbon width of 3.9$\pm$0.6 nm. In a close-up STM measurement of nanoribbon
as in Fig. 1d, parallel stripes with a period of $\sim$1.5 nm are identified, similar to the previous report\cite{CDing2019}. The stripe structure
determines the Q1D nature of electronic properties studied later. In Ref. 13, considerable quantity of clusters emerge in the
nanowire sample with nominal amount of 0.45 unit cell (0.45uc). Whereas here in Sample I, no cluster is observed (Fig. 1a), suggesting the
elevated coverage limit. By counting the area and orientations of all the nanoribbons in Fig. 1c, most of the nanoribbons are found to align
along two mutually perpendicular directions, i.e., the [100] and [010] axes of SrTiO$_3$ substrate (Fig. 1e).

To determine the in-plane lattice symmetry and lattice constant of nanoribbon, reflection high-energy electron diffraction (RHEED) measurements are performed
on Sample II. The RHEED patterns of two different geometries are shown in Fig. 1f-i. In Fig. 1f, the RHEED image only displays CoSb$_{1-x}$ streaks
with no noticeable superstructures, as evidenced by the single set of diffraction peaks in Fig. 1g. The absence of reconstruction of CoSb$_{1-x}$ streaks elucidates that the nanoribbon adopts in a lattice close to the tetragonal phase. When the glancing angle is increased and the incident beam is slightly
off the [100] direction (Fig. 1h), both the sharp diffraction spots of SrTiO$_3$ substrate and RHEED streaks of CoSb$_{1-x}$ are obtained. Based on the
distances between (00) and (02) diffraction peaks from CoSb$_{1-x}$ and SrTiO$_3$ substrate (Fig. 1i), the lattice constant of CoSb$_{1-x}$ is estimated
to be 4.0 \AA. Moreover, in Fig. 1i, there is no RHEED streak from the orthorhombic CoSb reported in Ref. 13, further demonstrating
that CoSb$_{1-x}$ nanoribbon possesses a different lattice structure from the orthorhombic phase\cite{CDing2019}.
Therefore, the corresponding unit cell should have the lattice constant of tetragonal phase $\sim$4.0 \AA ~along one axis, while have the period of the Q1D stripe $\sim$1.5 nm along the other axis.

\noindent \textbf{Experimental FS topology and band structures}

\noindent To unravel the electronic properties of CoSb$_{1-x}$ nanoribbons, we have scrutinized the electronic structure by ARPES experiments. The measured
FS of Sample III is shown in Fig. 2a, and the schematic FS is sketched in Fig. 2b. The Q1D character of FS topology can be better clarified
in Fig. 2c, where we plot the momentum distribution curves (MDCs) of FS mapping in Fig. 2a (indicated by translucent green rectangle in Fig. 2b).
Within experimental uncertainty, the FS forms straight line with no noticeable lateral dispersions, demonstrating the electrons are under Q1D
confinement. Moreover, since the experimental band structure is mostly a superposition of bands from mutually perpendicular nanoribbons,
the detected FS adopts two sets of Q1D features.
Based on the FS topologies, the Q1D Brillouin zone (BZ) size along the longitudinal direction of nanoribbon is consistent with the distance between diffraction streaks of nanoribbons in the RHEED measurements.

The near-$E_{\rm F}$ ARPES spectra of Sample III around $\Gamma$ and $X$ are measured along cuts\#1 and \#2 in Fig. 2b, respectively (Fig. 2d-g).
The hole-like $\alpha$ band and electron-like $\gamma$ band define the Q1D FSs. Although clusters occur in Sample III (Fig. 1c),
their contribution to the band dispersions is not detectable and the resolved bands are the same as that of Sample I and II without clusters
(see Supplementary Fig. 2).
Further, despite of the limited height of nanoribbons, quantum well states are not resolved in experiments, the identified
$\alpha$ and $\gamma$ are two different bands (see Supplementary Fig. 3 and Supplementary Note 1).
We carry out band calculations by confining monolayer CoSb to Q1D along [010] direction (Fig. 2h) while retaining the tetragonal crystal structure. Both the calculated FSs (Fig. 2i) and essential band structures (Fig. 2j) resemble our experimental observations. Specifically, the hole bands around $\Gamma$ correspond to the $\alpha$ band and the electron bands around $X$ correspond to the $\gamma$ band.
By further comparing the 2D calculations in Ref. 12 with our Q1D calculations (Fig. 2h-j), connections can be found between CoSb$_{1-x}$ nanoribbon and monolayer CoSb\cite{WDing2020}.
Therefore, the CoSb$_{1-x}$ nanoribbon system is likely a close counterpart of monolayer CoSb in reduced dimension.

By calculating the volumes of both the hole- and electron-like FSs relative to the Q1D BZ size in Fig. 2a, CoSb$_{1-x}$ is found to be slightly hole doped with 0.08 holes/Co based on the Luttinger volume and double degeneracy of the FSs\cite{WDing2020} (see more discussion in Supplementary Note 2).
This doping value is distinct from that of the heavily electron doped monolayer FeSe/SrTiO$_3$ (0.12 electrons/Fe)\cite{SHe2013,STan2013}, where the significant transfer of electrons from SrTiO$_3$ substrate to monolayer FeSe originates from the large work function difference between them\cite{ZhaoW2018,ZhangHM2017}.
Therefore, the charge transfer and work function difference in CoSb$_{1-x}$/SrTiO$_3$ interface should be opposite to that in monolayer FeSe/SrTiO$_3$, leading to the hole doping.
Further, because of the close connection between CoSb$_{1-x}$ nanoribbon and monolayer CoSb, it is reasonable to expect that the work function of monolayer CoSb is comparable to that of CoSb$_{1-x}$, and the charge transfer at an interface with SrTiO$_3$ substrate could be similar accordingly. This should serve as a useful piece of information in understanding the properties like potential superconductivity in CoSb-based interface in the future.

\noindent \textbf{TLL origin of the spectral weight suppression}

\noindent Notably, the ARPES spectral weight of CoSb$_{1-x}$ gradually decreases as approaching $E_{\rm F}$ (Fig. 2d-g and Supplementary Fig. 2). We examine the temperature evolution of energy distribution curves (EDCs) at $k_{\rm F}$ of Sample I (cut\#1 in Fig. 2b). As plotted in Fig. 3a, the spectral weight is continuously depleted towards $E_{\rm F}$, and there is no Fermi edge and well-defined quasiparticle peak. The suppression near $E_{\rm F}$ is clearer by comparing the $k$-integrated EDCs (cut\#1 in Fig. 2b) to a polycrystalline Au spectrum with a steep Fermi edge in Fig. 3b.
Although the nanoribbons misaligned from [100] and [010] orientations may give rise to ``polycrystallinity" and momentum-averaged spectra with reduced weight near $E_{\rm F}$, it would not deplete the density of states (DOS) at $E_{\rm F}$. As shown in Fig. 3c, similar spectral weight suppression is also observed in the STS spectra obtained on an individual nanoribbon. Therefore, the misaligned nanoribbons have minor effect on the spectral weight suppression
near $E_{\rm F}$.
This spectral weight depletion is not likely
due to the disorder effect either. As suggested by the Altshuler-Aronov theory\cite{AltshulerBL1979}, the DOS near $E_{\rm F}$ of disordered
metals is the summation of a finite constant and a power-law-like function\cite{KobayashiM2007}, which is distinct from our observation that the DOS nearly decreases to zero at $E_{\rm F}$. The similar momentum broadening of $\alpha$ band in Sample I and III indicates similar level of disorder scattering and similar angular distribution of nanoribbons (see Supplementary Fig. 4), further implying the origin of spectral weight suppression is intrinsic.
In a Peierls instability, the Q1D FS
topology would induce a large charge density wave (CDW) susceptibility, which may lead to a CDW order or sizable CDW fluctuations\cite{Gruner1994}.
The CDW order could be excluded in our case as the STS data in Fig. 3c exhibit continuous suppression towards $E_{\rm F}$ without any signature
of CDW gap formation, in contrast to the well-defined gap structure in the STS spectra of other typical Q1D CDW systems\cite{BarjaS2016,WangL2020}.
On the other hand, the CDW fluctuations could cause a pseudogap
opening, which may explain the spectral weight suppression\cite{PLee1973,RMcKenzie1996}. However, the suppressed spectral weight in a pseudogap would be
filled up following a linear function of the temperature\cite{KondoT2009}, which is not observed in our case (Fig. 3a).

The spectral weight suppression can be well accounted for by potential TLL state in CoSb$_{1-x}$. Serving as the spectroscopic fingerprints of
TLL, the power-law-like suppression and scaling behavior for low-energy DOS have been observed in several Q1D materials, like (TaSe$_4$)$_2$I\cite{Dardel1991}, Li$_{0.9}$Mo$_6$O$_{17}$\cite{WangF2006}, Au/Ge(001)\cite{Meyer2011}, Bi/InSb(001)\cite{Ohtsubo2015}, and K$_2$Cr$_3$As$_3$\cite{Watson2017}.
By a linear fit to the integrated spectra at 25 K in the double-logarithmic plot (Fig. 3d), the power-law-like manner of DOS can be straightforwardly
visualized. Because the finite temperature effect has not been considered, the curves at higher temperatures deviate from the simple power law below
the binding energy of $\sim$20 meV, which have also been observed in K$_2$Cr$_3$As$_3$\cite{Watson2017}.

To quantitatively model the DOS according to TLL theory with finite temperature effect included, we fit the integrated spectra based on the following
formula for photoemission spectral function\cite{Meden1993},
\begin{equation}\label{1}
I(\epsilon, T) \propto T^{\eta}\cosh\bigg(\frac{\epsilon}{2}\bigg)\ \bigg|\ \Gamma\ \bigg(\frac{1 + \eta}{2} + i\frac{\epsilon}{2\pi}\bigg)\
\bigg|^{2}f(\epsilon, T),
\end{equation}
where $\epsilon$ = $E$/$k_{\rm B}T$, $\Gamma$($\epsilon$) is the gamma function\cite{Gamma,Davis1959}, and $f$($\epsilon$, $T$) = ($e^{\epsilon}$ + 1)$^{-1}$
is the Fermi-Dirac distribution function. After convolved with Gaussian function of experimental energy resolution (${\Delta}E$ $\sim$ 9 meV), the
fittings well reproduce the experimental spectra at different temperatures (Fig. 3e). The analysis is only done in low-energy regime with binding
energy lower than 60 meV, since at higher energies the $\beta$ band could be involved in. The obtained power index $\eta$ shows a temperature
renormalization from 0.81 at 25 K to 0.76 at 70 K. A similar renormalization rule has been previously reported in Li$_{0.9}$Mo$_6$O$_{17}$, the
origin of which is proposed as the interacting charge neutral critical modes that emerge from the two-band nature of the material\cite{WangF2006}.
Here in CoSb$_{1-x}$, there are also two bands crossing $E_{\rm F}$ ($\alpha$ and $\gamma$), we thus suggest that the temperature renormalization
of $\eta$ might be understood in terms of that in Li$_{0.9}$Mo$_6$O$_{17}$.

Equation 1 further shows a universal power-law-like scaling relation, i.e., normalize the spectral intensity $I$($\epsilon$, $T$) at different
temperatures to $T^{\eta}$ and plot $I$/$T^{\eta}$ versus temperature-renormalized energy $\epsilon$ = $E$/$k_{\rm B}T$, the renormalized spectra
at different temperatures should coincide with each other if the scaling factor $\eta$ is chosen properly. As shown in Fig. 3f and
Supplementary Fig. 5,
within experimental uncertainty, the corresponding spectra can be well scaled when $\eta$ = 0.74$\pm$0.05. The best scaling behavior is illustrated by comparing the scaling
plots with various values of $\eta$ (see Supplementary Fig. 5), as $\eta$ departs from 0.74, the scaling becomes increasingly poor.
Besides, based on Eq. 1, the spectral intensity at $E_{\rm F}$ should manifest as $I$(0, $T$) $\propto$ $T^{\eta}$, showing the power-law-like
scaling as a function of temperature. As plotted in the inset of Fig. 3f, the temperature dependent $I$(0, $T$) follows a power-law-like manner with
$\eta$ $\sim$ 0.74, agreeing well with the above analysis.
Similar values of $\eta$ (0.75-0.80) have also been reproduced by fitting the STS\cite{Bockrath1999,Stuhler2020} and ARPES spectra from two different samples (black curves in Fig. 3c and Supplementary Fig. 6, respectively).
In the framework of TLL theory\cite{Braunecker2012}, $\eta$ is directly related to the TLL parameter $K$,
which characterizes the electron-electron correlation strength, via $\eta$ = ($K$ + $K^{-1}$ - 2)/4. We can deduce a value
of $K$ $\sim$ 0.21, indicating CoSb$_{1-x}$ is in the strongly correlated regime compared with $K$ = 1 for the noninteracting Fermi liquid regime.
The parameter is also comparable with that of other Q1D systems, for example, single-wall carbon nanotubes ($K$ $\sim$ 0.25)\cite{Bockrath1999},
MoSe$_2$ mirror twin boundaries ($K$ $\sim$ 0.28)\cite{XiaYP2020}, SrNbO$_{3.41}$ single crystals ($K$ $\sim$ 0.2)\cite{Campos2010}, and
Li$_2$Mo$_6$Se$_6$ nanowires ($K$ $\sim$ 0.15)\cite{Venkataraman2006}.
The strong correlation behavior in the Q1D CoSb$_{1-x}$ is the combined effect of the dimensional confinement and the electron-electron interaction in CoSb itself.
As shown by our calculation in Fig. 2j, the overall bandwidth of Q1D CoSb is reduced by approximately half from that of monolayer tetragonal CoSb\cite{WDing2020}, corresponding to a moderate level of correlation enhancement by the dimensional confinement. Another effect of the dimensional confinement is eliminating the quasiparticle weight by inducing the TLL state, which is confirmed by our observation. The correlation strength of CoSb itself is hard to quantify due to the lack of quasiparticle, while the observed dispersion in ARPES is from incoherent features that follow the bare dispersion without correlation effect. Indeed, the dispersions of $\alpha$ and $\gamma$ bands extracted from MDCs match those in calculations (Fig. 2j). It calls for future experiments on monolayer CoSb films to reveal the underlying correlation strength.

\section*{DISCUSSION}

It is worth noting that the Coulomb blockade from the size effect has been observed on isolated CoSb-related nanowires\cite{CDing2019}.
Statistics of the topography in Fig. 1a gives the total length of 38$\pm$6 nm for connected nanoribbon domains, corresponding to an average
gap size of $\sim$3-5 meV from the Coulomb blockade studied by STS\cite{CDing2019}. From this perspective, it is only $\sim$10\% of the leading
edge position (binding energy of $\sim$40 meV) of spectral weight suppression in Sample I (Fig. 3a), indicating the Coulomb blockade would only
have a minor contribution to the spectral weight suppression. Nevertheless, the Coulomb blockade has been observed by photoemission measurements
in several interfacial materials, and been suggested as the dynamic final state effect during the photoemission process, which would induce
statistical spectra broadening in the energy shifts\cite{Starowicz2002,HovelH2004}. Therefore, the Coulomb blockade may have a decoration effect
on the spectral shape of Sample I, making a contribution to the spectral weight suppression near $E_{\rm F}$, while the good agreement of our data
with the power-law analysis indicates the dominant role of the TLL state in CoSb$_{1-x}$ nanoribbons.

In summary, we have fabricated CoSb$_{1-x}$ nanoribbons with Q1D stripes on SrTiO$_3$(001) substrates and investigated the electronic structure.
The Q1D FS topology is revealed. CoSb$_{1-x}$/SrTiO$_3$ is slightly hole doped, where the interfacial charge transfer is opposite to that
in monolayer FeSe/SrTiO$_3$. The spectral weight near $E_{\rm F}$ shows power-law-like suppression and obeys a universal temperature scaling,
suggesting CoSb$_{1-x}$ belongs to the TLL. The corresponding TLL parameter of $\sim$0.21 uncovers its strongly correlated nature. Our study
not only demonstrates a canonical TLL system, but also sheds light on the understanding of the physics in CoSb-based interface.

\section*{METHODS}
\textbf{Synthesis of nanoribbons}

\noindent CoSb$_{1-x}$ nanoribbons are fabricated by coevaporating Co (99.995\%) and Sb (99.999\%) from Knudsen cells on Nb-doped (0.5 wt\%) SrTiO$_3$(001)
substrates. During the growth of samples with nominal amounts of 0.45uc, 0.55uc, and 1uc, which are denoted as Sample I-III, the substrates are
kept at $\sim$130-160 $^{\circ}$C, the cell temperatures of Co and Sb are set at 1162 and 337 $^{\circ}$C, respectively. The corresponding growth
rate is 0.02 uc/min.

\noindent \textbf{ARPES and STM/STS measurements}

\noindent ARPES experiments are performed $\emph{in situ}$ equipped with Fermi Instruments helium discharge lamp (21.22 eV He-I$\alpha$
light) and a Scienta R4000 analyzer. The energy and angular resolutions are set to 9 meV and 0.3$^{\circ}$, respectively. During the measurements,
the temperature is kept at 40 K if not specified otherwise and the working vacuum is maintained better than 4 $\times$ 10$^{-11}$ mbar. The samples
are then transferred to the RHK (data in Fig. 1a-d) and CreaTec (data in Fig. 3c) STM systems via a vacuum suitcase with base pressure of 3 $\times$ 10$^{-10}$ mbar. STM/STS measurements are conducted at low temperatures with Pt/Ir tips.

\noindent \textbf{Computational methods and details}

\noindent The energetic calculations and geometrical relaxations are performed using the Vienna \emph{ab initio} Simulation Package\cite{KresseG1996} within density functional theory. Valence electrons are described using the projector-augmented-wave method\cite{Blochl1994,KresseG1999}. The exchange and correlation functional are treated using the Perdew-Bruke-Ernzerhof\cite{Perdew1996} parametrization of generalized gradient approximation. The kinetic energy cutoff of the plane-wave basis is chosen to be 280 eV. A $\Gamma$-centered 15$\times$15$\times$1 Monkhorst-Pack\cite{Methfessel1989} $k$-mesh is used for BZ sampling, and a much denser $k$-mesh of 45$\times$45$\times$1 is used for accurate DOS calculations. Electronic minimizations are performed with an energy tolerance of 10$^{-6}$ eV, and optimized atomic structures are achieved when forces on all the atoms are  $\textless$ 0.01 eV/\AA. To model the Q1D CoSb nanoribbons, the supercells contain a ribbon of the freestanding tetragonal CoSb with width of 3uc ($\sim$1.2 nm), and to avoid the interference between neighboring nanoribbons, the ribbons are separated by a 2uc-wide empty space ($\sim$0.78 nm) laterally on each side under periodic boundary conditions.
To model a freestanding 2D monolayer CoSb, the supercells contain a slab of the monolayer CoSb with a vacuum region of more than 15 \AA. The spin-orbit coupling effect is considered in band structure and DOS calculations.

\section*{CODE AVAILABILITY}
The computer codes used for the band structure calculations in this study are available from the corresponding authors upon reasonable request.

\section*{DATA AVAILABILITY}
The data that support the findings of this study are available from the corresponding authors upon reasonable request.

\section*{ACKNOWLEDGMENTS}
\noindent This work was supported by the National Natural Science Foundation of China (Grant Nos. 11790312, 12074074, and 11904144), the National Key R\&D
Program of the MOST of China (Grant Nos. 2017YFA0303004 and 2016YFA0300200), and Shanghai Municipal Science and Technology Major Project (Grant No. 2019SHZDZX01).

\section*{COMPETING INTERESTS}
The authors declare no competing interests.

\section*{AUTHOR CONTRIBUTIONS}
D.L.F., Z.Y.Z., H.C.X., and R.P. conceived the projects and experiments.
R.L. and M.Y.N.L. synthesized the samples using MBE.
R.L. and M.Y.N.L. performed ARPES measurements with the assistance of X.Y.C. and X.P.S.
R.L. and M.Y.N.L. performed STM measurements with the assistance of W.T.Y., R.T., S.Y.D., Y.J.Y., and T.Z.
W.J.D., P.C., and Z.Y.Z. performed band structure calculations.
R.L., H.C.X., R.P., T.Z., and D.L.F. analysed the experimental data.
R.L. plotted the figures.
R.L., H.C.X., and D.L.F. wrote the manuscript with input from all the authors.

\section*{ADDITIONAL INFORMATION}
\textbf{Supplementary information} accompanies the paper on the $\textit{npj Quantum Materials}$ website.

\noindent \textbf{Correspondence} and requests for materials should be addressed to H.C.X. and D.L.F.

\newpage

\begin{figure}
  \begin{center}
  \includegraphics[width=1.0\columnwidth]{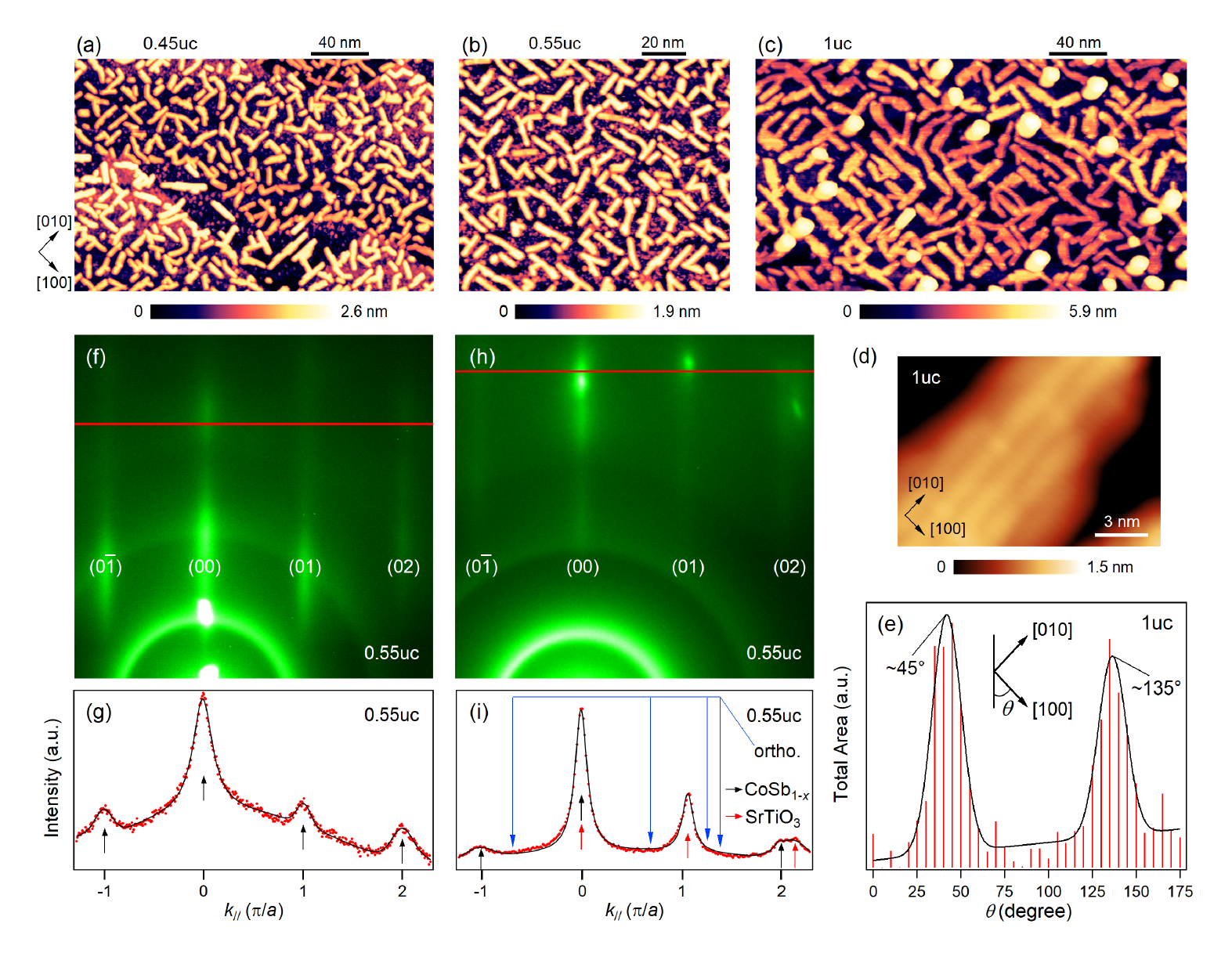}
  \end{center}
  \caption{\textbf{Surface morphology and in-plane lattice symmetry of CoSb$_{1-x}$ nanoribbons.}
  (a)-(c) STM topographies of Sample I (0.45uc), II (0.55uc), and III (1uc), respectively, measured with $V_{\rm s}$ = 3 V, $I$ = 30 pA. Clusters
  of $\sim$4 nm in height form on top of Sample III.
  (d) STM image of a [010] oriented CoSb$_{1-x}$ nanoribbon of Sample III ($V_{\rm s}$ = 3 V, $I$ = 30 pA).
  (e) Counted area and orientations of all the nanoribbons in (c). The definition of $\theta$ is shown in the inset. The black curve is the fitting
  result by two Gaussian peaks and a linear background.
  (f),(g) RHEED image of Sample II with incident beam along the [100] direction and corresponding RHEED intensity curve along the red cut, respectively.
  (h),(i) Same as (f),(g) with a larger glancing angle and incident beam slightly off the [100] direction.
  In (g),(i), the black curves are the fitting results by Gaussian peaks and quadratic polynomial backgrounds, the black and red arrows indicate the
  diffraction signals from CoSb$_{1-x}$ and SrTiO$_3$ substrate, respectively. $a$(= 4.0 \AA) is the lattice constant of CoSb$_{1-x}$. The blue arrows
  represent the anticipated streak positions of orthorhombic CoSb phase in Ref. 13.
  }
\end{figure}

\begin{figure}
  \begin{center}
  \includegraphics[width=1.0\columnwidth]{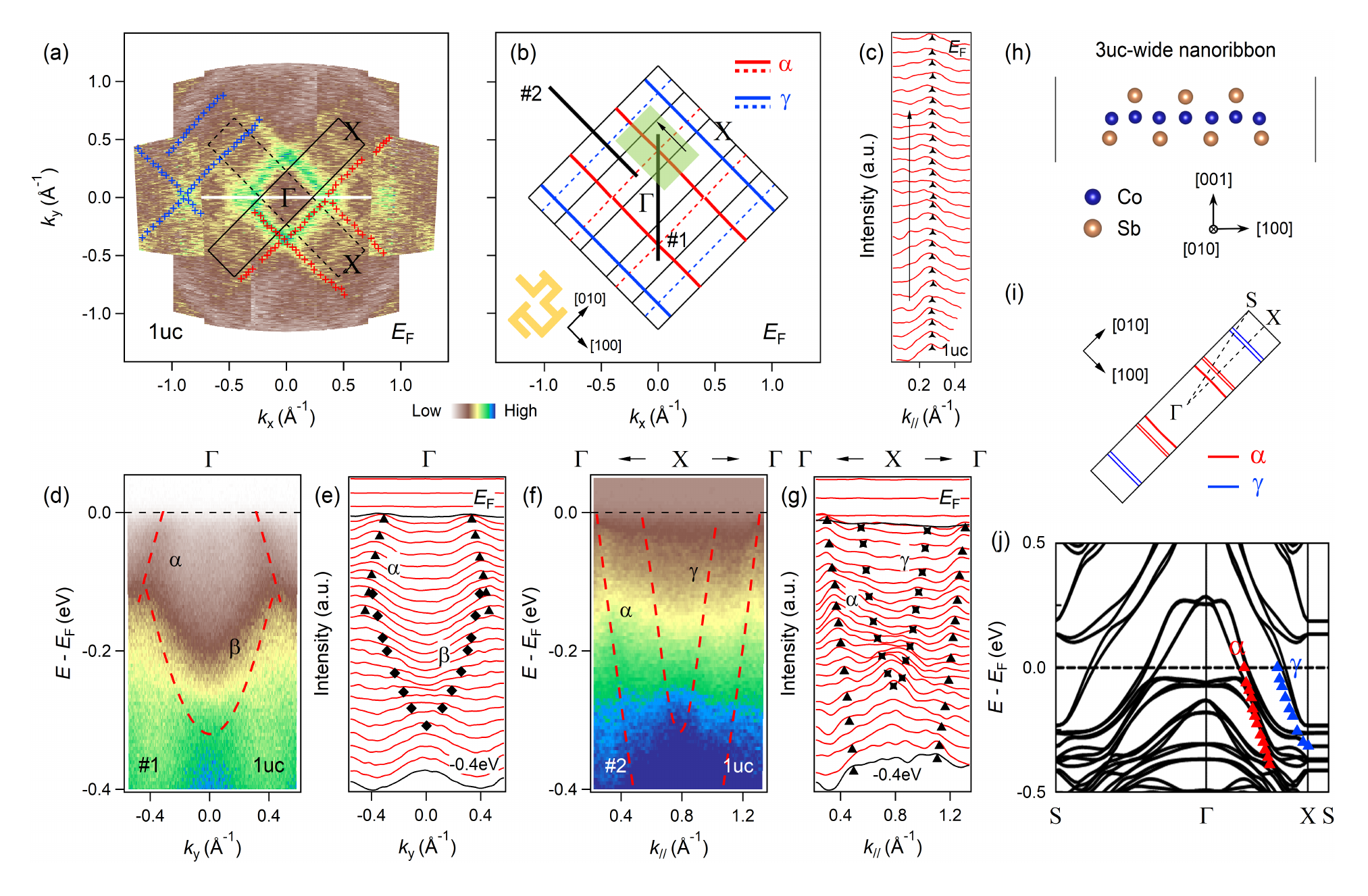}
  \end{center}
  \caption{\textbf{FS topology and band structures of CoSb$_{1-x}$ nanoribbons.}
  (a) Constant-energy ARPES image of Sample III (1uc) obtained by integrating the photoemission intensity within $E_{\rm F}$$\pm$40 meV. The red and
  blue cross marks are measured Fermi crossings.
  (b) Schematic FS topology of (a).
  Inset: Sketches of the [100] and [010] oriented CoSb$_{1-x}$ nanoribbons.
  The black solid and dashed lines in (a) represent the Q1D BZs of [010] and [100] oriented nanoribbons, respectively. The black solid
  lines in (b) indicate a series of Q1D BZs of [010] oriented nanoribbons, while the BZs of [100] orientation are omitted for better visualization. The solid and dashed schematic FS lines in (b) represent that of [010] and [100] oriented nanoribbons, respectively.
  (c) MDC plot of the constant-energy spectra in (a), which is indicated by the translucent green area in (b). The triangle markers are extracted peak positions.
  (d),(e) Intensity plot and corresponding MDCs along cut\#1 in (b), respectively.
  (f),(g) Same as (d),(e) along cut\#2 in (b). The red dashed curves in (d),(f) are guides to the eye for the experimental band dispersions. The black
  markers in (e),(g) are extracted peak positions.
  (h) Supercell contains a [010] oriented ribbon of the freestanding tetragonal CoSb with width of 3uc ($\sim$1.2 nm).
  (i) Calculated FSs of the CoSb supercell in (h). The BZ of the supercell represented by the black solid lines shows an aspect ratio of 5: 1.
  (j) Calculated band structures of 3uc-wide CoSb nanoribbon. The red and blue markers are experimental $\alpha$ and $\gamma$ bands extracted from the MDCs
  in (g), respectively.
  }
\end{figure}

\begin{figure}
  \begin{center}
  \includegraphics[width=0.85\columnwidth]{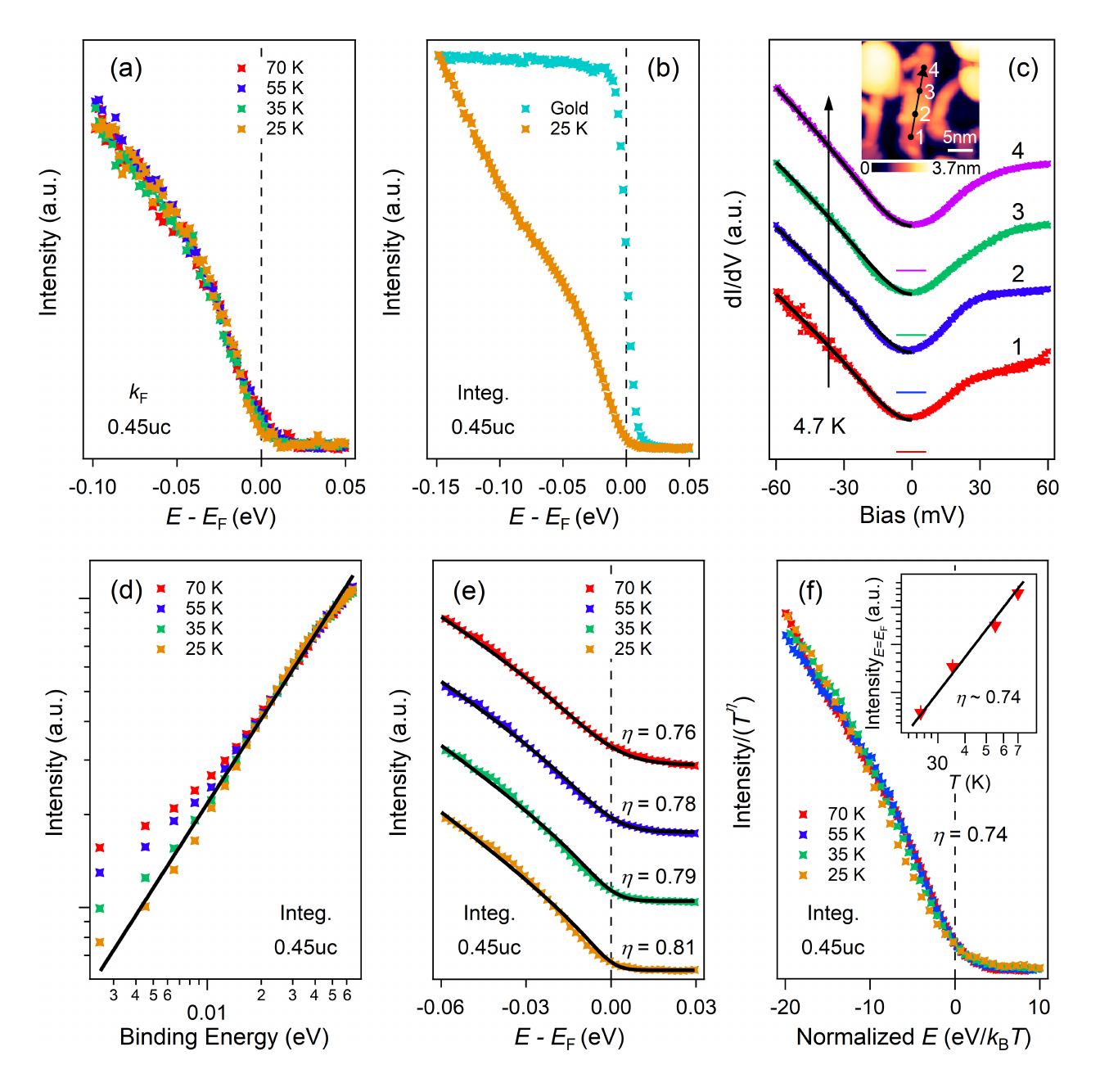}
  \end{center}
  \caption{\textbf{Evidence of TLL state in CoSb$_{1-x}$ nanoribbons.}
  (a) Temperature dependent EDCs taken at $k_{\rm F}$ (cut\#1 in Fig. 2b) of Sample I (0.45uc).
  (b) Integrated spectra (cut\#1 in Fig. 2b) at $T$ = 25 K in the range of -0.56 $\textless$ $k_y$ $\textless$ 0.56 \AA$^{-1}$, compared
  to a polycrystalline Au spectrum.
  (c) $dI$/$dV$ spectra taken along the black arrow in the inset on an individual nanoribbon, with $V_{\rm s}$ = 60 mV, $I$ = 100 pA
  ($T$ = 4.7 K, $\Delta$$E$ $\sim$ 2 mV). For clarity, each spectrum is offset with a constant conductance and the zero is indicated by a horizontal bar.
  The black curves are the fittings to TLL model. The asymmetric amplitude of DOS may originate from the local electronic states affected by the size
  effect of nanoribbons.
  Inset: STM topography of CoSb$_{1-x}$ measured with $V_{\rm s}$ = 0.8 V, $I$ = 30 pA.
  (d) Logarithmic plot of the temperature dependent integrated spectra [same as (b)]. The black line is a linear fit of data at 25 K in the
  binding energy range of 3-60 meV.
  (e) Same as (d) plotted in a linear scale. The spectra are vertically shifted for clarity. The black curves are the fittings to Eq. 1,
  convolved with Gaussian function of instrumental energy resolution (${\Delta}E$ $\sim$ 9 meV).
  (f) Universal scaling plot of the spectra in (e), where the spectral intensity is scaled as $I$/$T^{0.74}$, and energy is normalized to
  $E$/$k_{\rm B}T$.
  Inset: Logarithmic plot of the temperature dependent spectral intensity at $E_{\rm F}$, where the data are extracted from (e).
  The black line is a linear fit, implying the power-law-like behavior with power index of $\sim$0.74.
  }
\end{figure}


\begin{thebibliography}{1}
\bibitem{Gamma}
The gamma function, which is known as the Euler's integral of the second kind, is one commonly used extension of the factorial function to
complex numbers. See Ref. 37 for detailed definitions.
\end{thebibliography}
\end{document}